\documentclass[journal,transmag]{IEEEtran}

\ifCLASSINFOpdf
  
\else
\fi

\usepackage{graphicx}
\usepackage{caption}
\usepackage{amsmath}

\begin{document}

\title{Design, Engineering and Optimization of a Grid-Tie Multicell Inverter for Energy Storage Applications}


\author{\IEEEauthorblockN{A. Ashraf Gandomi \IEEEauthorrefmark{1},
S. Saeidabadi \IEEEauthorrefmark{1},
M. Sabahi \IEEEauthorrefmark{1}, 
M. Babazadeh \IEEEauthorrefmark{1}, and
Y. Ashraf Gandomi \IEEEauthorrefmark{2}}
\IEEEauthorblockA{\IEEEauthorrefmark{1}Department of Electrical and Computer Engineering,\\ University of Tabriz, Tabriz}
\IEEEauthorblockA{\IEEEauthorrefmark{2}Electrochemical Energy Storage and Conversion Laboratory,\\ Department of Mechanical, Aerospace and Biomedical Engineering,\\ University of Tennessee, Knoxville, Tennessee, USA}

\thanks{Manuscript submitted August 26, 2017.
Corresponding author: A. Ashraf Gandomi (email: a.ashrafgandomi91@ms.tabrizu.ac.ir).}}

\markboth{}
{Shell \MakeLowercase{\textit{et al.}}: Bare Demo of IEEEtran.cls for IEEE Transactions on Magnetics Journals}

\IEEEtitleabstractindextext{%
\begin{abstract}
Multilevel converters have found many applications within renewable energy systems thanks to their unique capability of generating multiple voltage levels. However, these converters need multiple DC sources and the voltage balancing over capacitors for these systems is cumbersome. In this work, a new grid-tie multicell inverter with high level of safety has been designed, engineered and optimized for integrating energy storage devices to the electric grid. The multilevel converter proposed in this work is capable of maintaining the flying capacitors voltage in the desired value. The solar cells are the primary energy sources for proposed inverter where the maximum power density is obtained. Finally, the performance of the inverter and its control method simulated using PSCAD/EMTDC software package and good agreement achieved with experimental data.
\end{abstract}

}

\maketitle

\IEEEdisplaynontitleabstractindextext

\IEEEpeerreviewmaketitle

\section{Introduction}

\IEEEPARstart{T}{he} application of renewable energy sources (i.e. solar wind energy) into an electric grid requires high performance energy storage devices along with various types of power electronics (i.e. rectifiers, converters and inverters). 
Figure 1 includes the schematic of a hybrid energy storage system in which a renewable energy source (here photovoltaic modules) along with an energy storage device has been implemented to the electric grid via the utilization the multilevel inverters \cite{gandomi2015dc,saeidabadi2017novel,gandomi2017maximum}.

\begin{figure}[!t]
\centering
\captionsetup{justification=centering}
\includegraphics[width=3.5in]{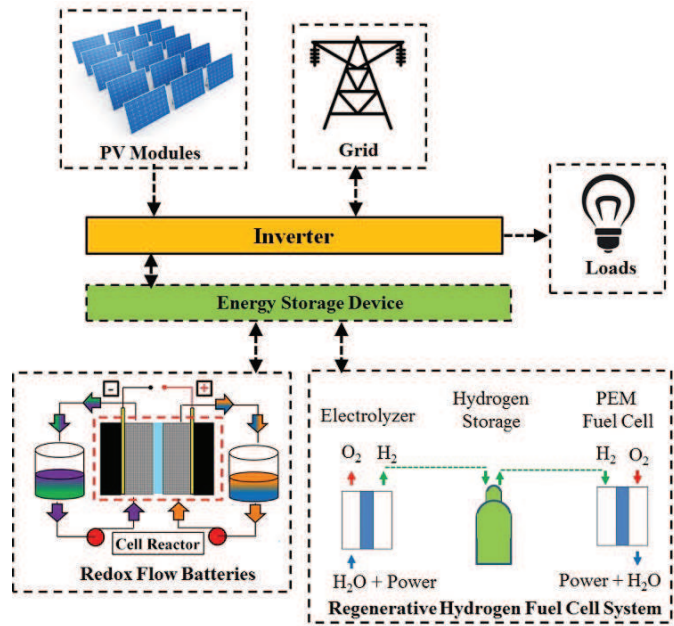}
\caption{Schematic of a hybrid energy system.}
\label{fig_sim}
\end{figure}

As shown in Fig. 1, battery-based devices and hydrogen-based energy storage technologies are promising. A good review on the battery- and hydrogen-based energy storage has been provided recently \cite{pellow2015hydrogen}. For hydrogen based energy storage, an electrolyzer along with a polymer electrolyte membrane (PEM) fuel cells are utilized. Although being very promising, the water management for electrolyzers and PEM fuel cells remains a challenging issue to be addressed \cite{gandomi2013assessing,gandomi2016water}. Also, the hydrogen gas storage technologies and infrastructure for regenerative hydrogen fuel cell systems is required to be addressed appropriately \cite{zehsaz2011plastic,zehsaz2012bifurcation}. Several battery technologies are developed for energy storage technologies that redox flow batteries are among the most promising devices . However engineering the redox flow batteries for optimum operation requires detailed mathematical models \cite{gandomi2014concentrated,yousefzadi2016anisotropic,gandomi2016situ} and experimental diagnostics \cite{gandomi2016coupled,ashraf2017influence}.

The application of multilevel converters for the renewable energy systems has been increased recently \cite{saeidabadi2017two,hosseini2015attempt,gandomi2015control,gandomi2015transformer}. This is mostly because of high energy output (in the range of MW), high efficiency and low electromagnetic interference being offered via their application \cite{varesi2017improved,gandomi2017high,saeidabadi2017new}.

Three-level converter first proposed by Nabae \cite{nabae1981new} and the converters with increased number of levels developed later to obtain higher levels of voltage and reduce the harmonic content of the output voltage. However, increasing the number of voltage levels results in increased system complexity and severe voltage balancing problems. Different topologies for these types of converters have been utilized including Neutral Point Clamped (NPC) and variant types of multicell inverters \cite{sadigh2013reduced,rodriguez2010survey,banaei2013series}. The NPC topology suffers from diode clamped and capacitor voltage balancing issues where the multicell inverters do not suffer from the same issue. Therefore, multicell inverters are promising candidates to generate increased voltage levels for high power generation applications. The multicell inverters have different types including the Cascaded Multicell (CM), Flying Capacitor Multicell (FCM) and Stacked Multicell (SM) inverters \cite{mcgrath2007natural,lezana2009fault,meynard1997modeling}. The FCM inverters along with the SM inverters have many unique advantages for medium voltage applications such as operation without transformer and naturally capacitor voltage balancing ability. A comparison of the different components being used with these different inverters for generating   levels had been provided in Table. I.

Table. I. comparison of different inverter’s components

\begin{table}[!t]
\renewcommand{\arraystretch}{1.3}
\caption{Comparison of different inverter’s components}
\label{table_example}
\centering
\begin{tabular}{|c||c||c||c||c|}
\hline
Type & \# DC Sources & Magnitude & \# Switches & \# Capacitors\\
\hline
CM & $n$ & $E/n$ & $4n$ & $0$\\
\hline
FCM & $2$ & $E$ & $4n$ & $2n-1$\\
\hline
SM & $2$ & $E$ & $4n$ & $2(n-1)$\\
\hline
DFCM & $1$ & $E$ & $2n+2$ & $n-1$\\
\hline
\end{tabular}
\end{table}

The advantages of DFCM compared to conventional structures are eliminating the common point of DC sources, reducing the number of input DC sources, switches and capacitors. However, this structure is an island and in the grid-tie form of this inverter with presented control method, some difficulties arise including the power transfer from grid to input source and the possibility of discharging flying capacitors in case of connecting to the Photovoltaic (PV) cells.
In this paper a new topology for grid-tie multicell inverter has been developed. The proposed inverter has been designed based upon the DFCM inverter and accordingly shares the common advantages associated with the DFCMs including the elimination of common point of DC source, reduced number of DC sources and increased number of voltage levels. Also, the new design assures higher level of safety and via its utilization, the common problems related to the DFCMs has been eliminated (while extracting the maximum power from PV cells).

\section{PROPOSED MULTILEVEL INVERTER}

The proposed inverter is shown in Fig. 2. In this structure independent of the number of inverter cells and generated voltage levels, only the switches $S_n$ and $P_n$ in $cell_n$ are bidirectional. The oscillations of the PV cells voltage is mostly caused by climatic oscillations. In case of low PV cells and because of the cascaded diode of PV cells would not be functional and therefore, the possibility of discharging flying capacitors or grid voltage on the PV cells increases. Discharging the grid in the positive half cycle, the negative half cycle and discharging the flying capacitors are shown in Fig. 3. According to Fig. 3, in the proposed inverter, by using only two bidirectional switches, there would be no chance of discharging the grid and the flying capacitors on the PV cells. 

Table II, includes the switching states within the proposed inverter for generating 5-level output voltage. In this table, when the switch is on, the corresponding state number is one and for the off position it is zero. Also X in this table due to modes of charging or discharging of flying capacitor $C_1$, is either one or zero. Based on this table, in the states 2 and 4, charging or discharging of $C_1$  happens naturally, while in the other states, due to value of X, $C_1$ can be charged. The voltage balancing algorithm of $C_1$ in positive half cycle is shown in Fig. 3. Based on this figure, if the measured value of voltage of capacitor $C_1$ is lower than its set value ($v_{C1}$), $P_2$ (or $S_2$) will be on in the state 1 (or 3). This can continues until the voltage of $C_1$ reaches its specified value then $P_2$ will be off. 

\begin{figure}[!t]
\centering
\captionsetup{justification=centering}
\includegraphics[width=3.5in]{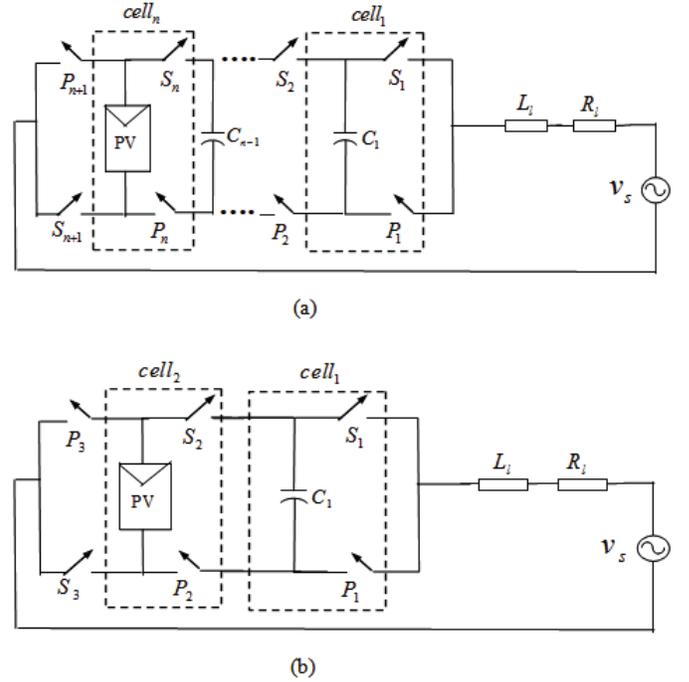}
\caption{(a) Proposed grid-tie multicell inverter and (b) Proposed 5-level grid-tie multicell inverter.}
\label{fig_sim}
\end{figure}

\begin{figure}[!t]
\centering
\captionsetup{justification=centering}
\includegraphics[width=3.5in]{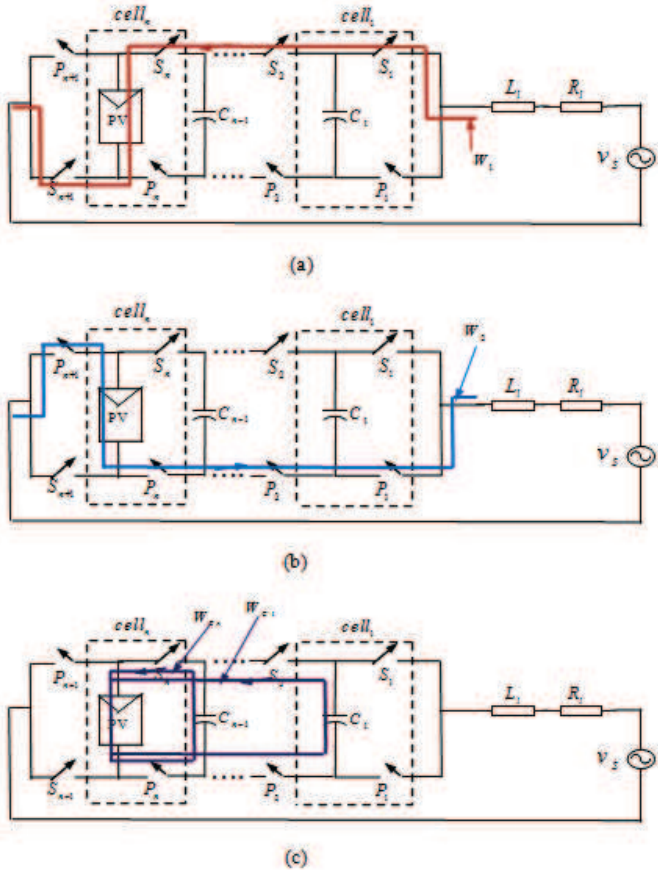}
\caption{(a) Discharging the grid in the positive half cycle, (b) Discharging the grid in the negative half cycle and (c) Discharging the flying capacitors.}
\label{fig_sim}
\end{figure}

\begin{table}[!t]
\renewcommand{\arraystretch}{1.3}
\caption{Switching states in the proposed inverter}
\label{table_example}
\centering
\begin{tabular}{|c||c||c||c||c||c||c|}
\hline
States & $S_1$ & $S_2$ & $S_3$ & $P_1$ & $P_2$ & $P_3$\\
\hline
1 & $1$ & $1$ & $1$ & $0$ & $X$ & $0$\\
\hline
2 & $1$ & $0$ & $1$ & $0$ & $1$ & $0$\\
\hline
3 & $1$ & $1$ & $0$ & $0$ & $X$ & $1$\\
\hline
4 & $0$ & $1$ & $0$ & $1$ & $0$ & $1$\\
\hline
5 & $0$ & $X$ & $0$ & $1$ & $1$ & $1$\\
\hline
\end{tabular}
\end{table}

\begin{figure}[!t]
\centering
\captionsetup{justification=centering}
\includegraphics[width=3.5in]{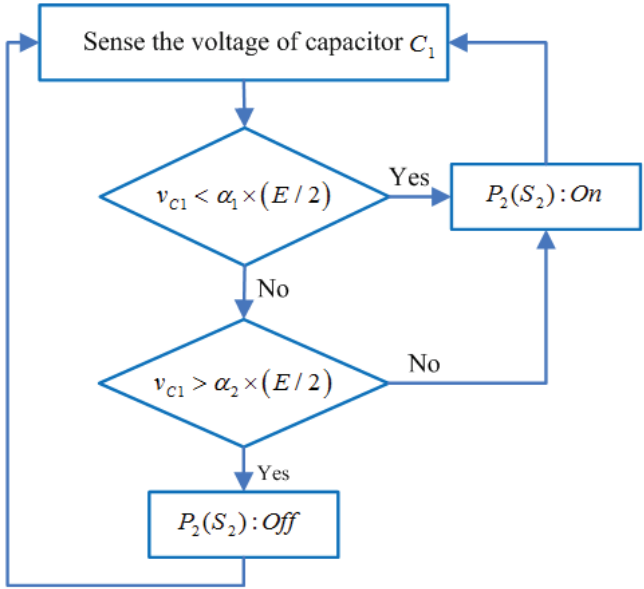}
\caption{Algorithm of charge and discharge of capacitor   in the positive half cycle of output voltage.}
\label{fig_sim}
\end{figure}

\section{PROPOSED CONTROL METHOD}

In the control method, it is assumed that the inverter output current tracks the favorite sinusoidal reference current. Also as it is shown in Fig. 5, it is assumed that the proposed inverter (base on IEEE 1547 standard) does not exchange any reactive power with the grid and only active power exchanges, therefore:

\begin{figure}[!t]
\centering
\captionsetup{justification=centering}
\includegraphics[width=3.5in]{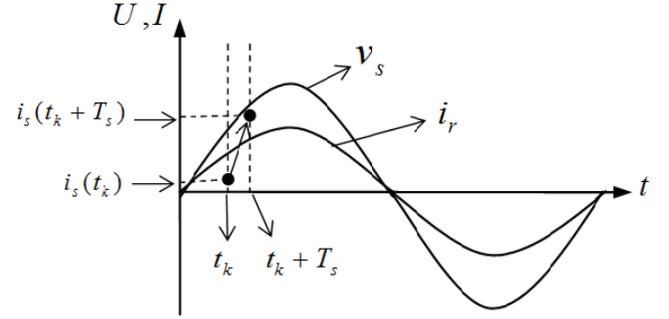}
\caption{Proposed control method.}
\label{fig_sim}
\end{figure}

\begin{equation}
v_s(t)=V_m\sin{(wt)}
\end{equation}

\begin{equation}
i_r(t)=I_m.\sin{(wt)}
\end{equation}

Figure 6 shows five output voltage levels that have been considered for the proposed inverter. The proposed control method should be capable of injecting power from inverter to grid at different operating conditions and therefore, the maximum voltage generated through the inverter ($E_1$) should be greater than the grid’s maximum voltage ($V_m$).

\begin{figure}[!t]
\centering
\captionsetup{justification=centering}
\includegraphics[width=3.5in]{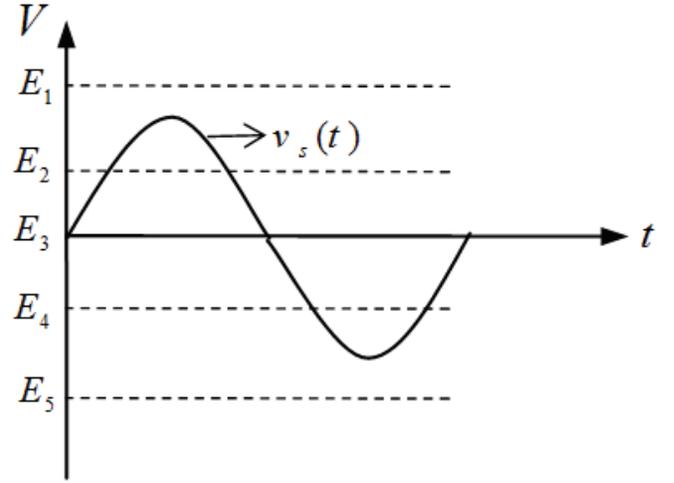}
\caption{The output voltage levels of proposed inverter.}
\label{fig_sim}
\end{figure}

Due to generating output voltage with low Total Harmonic Distortion (THD), the values of output voltages are considered as follows:

\begin{equation}
E_2=\frac{E_1}{2},E_3=0,E_4=\frac{-E_1}{2},E_5=-E_1
\end{equation}

The grid voltage and reference current is calculated according to (4) and (5) assuming the circuit current in the constant $t_k$ is $i_s(t_k)$:

\begin{equation}
v_s(t_k)=V_m\sin{(wt_k)}
\end{equation}

\begin{equation}
i_r(t_k)=I_m\sin{(wt_k)}
\end{equation}

The equivalent circuits of the proposed inverter are shown in Fig. 7.

\begin{figure}[!t]
\centering
\captionsetup{justification=centering}
\includegraphics[width=3.5in]{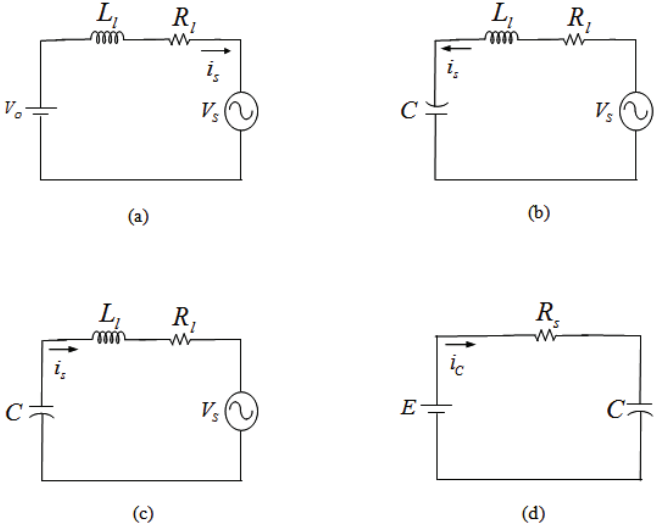}
\caption{(a) Equivalent circuit of the system, (b) Equivalent circuit of capacitor discharging state, (c) Equivalent circuit of capacitor discharging state and (d) Equivalent circuit of capacitor charging.}
\label{fig_sim}
\end{figure}

In Fig. 7, $R_l$ and $L_l$ are the line resistance and inductance between inverter and grid and $V_0$. According to Fig. 7(a):

\begin{equation}
L_1\frac{di_s(t)}{dt}+R_li_s(t)=V_0+v_s(t)=V_0+V_m\sin{(wt)}
\end{equation}

Then:

\begin{equation}
i_s(t_k)=I_k
\end{equation}

\begin{equation}
i_s(t)=K_1e^\frac{-R_lt}{L_l}+K_2\sin{(wt)}+K_3\cos{(wt)}+\frac{V_0}{R_l}
\end{equation}

\begin{equation}
K_1=\frac{I_k-\frac{V_m}{R_l^3+L_l^2w^2R_l}+\frac{V_mL_lw}{R_l^2+L_l^2w^2}}{e^\frac{-R_lt}{L_l}}
\end{equation}

\begin{equation}
K_2=\frac{V_m}{R_l^3+L_l^2w^2R_l}
\end{equation}

\begin{equation}
K_3=\frac{-V_mL_lw}{R_l^2+L_l^2w^2}
\end{equation}

In constant $t_k$, $i_s(t_k)$ is measured. Control system using (8), calculates five current for a five-level output voltage of inverter during the next switching period. Therefore, for five output voltage levels, five cost functions can be written as follow:

\begin{equation}
\Delta{I}=|i_r)t_k+T_s)-i_s(t_k+T_s)|
\end{equation}

From (12), proper switching state and consequently proper output voltage can be applied.
Initially, when the inverter connects to the network, the initial voltage of the capacitor is zero. At this point, no impulse current is drawn until the capacitor voltage reaches the proper value where the grid current is bypassed by $S_1$, $S_2$ and $P_3$ and simultaneously input source charged the capacitor by $S_2$ and $P_2$. When the capacitor voltage reaches its proper value, control method will follow their normal operation.
In Fig. 8(a) and (b), upper waveforms are inverter output voltage and the lower waveforms are the inverter output current. Based on Fig. 8(a), if the initial control method does not apply, 60\% of the original period (0.012 seconds) would be required for inverter to reach its own stable performance. This situation also results in over-drawn current as high as four times of the maximum nominal current through the circuit, which potentially damages the whole circuit. However, this situation has improved in Fig. 8(b) where the response time has decreased to about 0.0005 seconds. Therefore, the inverter only requires approximately 2.5\% of the original period to reach its own stable performance. Also in this case, the drawn current is within the safe range.

\begin{figure}[!t]
\centering
\captionsetup{justification=centering}
\includegraphics[width=3.5in]{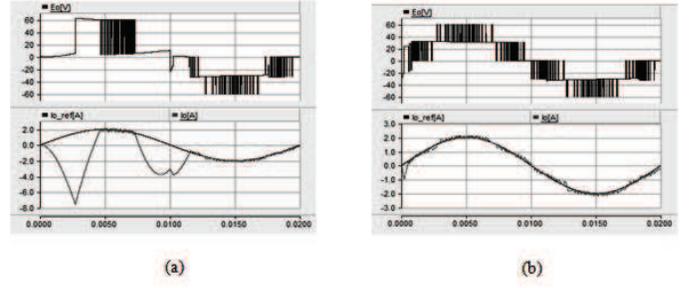}
\caption{Voltage stability (a) without initial control method and (b) with initial control method.}
\label{fig_sim}
\end{figure}

\section{FLYING CAPACITOR VOLTAGE BALANCING METHOD}
In this section, the method of capacitor voltage balancing is investigated. Based on Table II, the capacitor is discharged in the duration that output voltage equals to $E/2$ with positive current of circuit and output voltage equal to $-E/2$ with negative current of circuit. Based on the equivalent circuit of capacitor discharging states (Fig. 7(b) and (c)):

\begin{equation}
v_s(t)=V_m\sin{(wt)}
\end{equation}

\begin{equation}
i_s(t)=I_m\sin{(wt)}
\end{equation}

\begin{equation}
v_s(t)=V_c(t)+L_l\frac{di_L(t)}{dt}+R_li_s(t)
\end{equation}

\begin{multline}
v_c(t)=\sqrt{(V_m-R_lI_m)^2+(L_lI_m)^2}\sin[wt+ \\ 
arcsin(\frac{-L_lI_m}{\sqrt{(V_m-R_lI_m)^2+(L_lI_m)^2}})]
\end{multline}

In the equations formulated earlier ((13) to (16)),   is the grid voltage and $v_c(t)$ is the capacitor voltage. Also $i_s(t)$ is the circuit current that should be the sine wave because of tracking the reference sine wave.
According to Table II, unlike the case of the capacitor discharging state that was possible only in two operating modes, charging of capacitor happens in all five operating modes. When the inverter output voltage is equal to $E/2$ with negative current of circuit and equal to $-E/2$  with positive current of circuit, capacitor is charged naturally. But in the duration that output voltage is equal to $E$ or zero, charging of capacitor is controlled by switch $P_2$ and within the range that output voltage is equal to $-E$ or zero, charging of capacitor is controlled by switch $S_2$. The capacitor charging circuit in all five modes is shown in Fig. 7(d). In this figure, $R_s$ is the resistance of switches per current loop, therefore the following equations for a capacitor charging state can is obtained:

\begin{equation}
E=R_si_c(t)+v_c(t)
\end{equation}

\begin{equation}
i_c(t)=c\frac{dv_c(t)}{dt}
\end{equation}

\begin{equation}
v_c(t_1)=V_1
\end{equation}

\begin{equation}
v_c(t)=E+\frac{V_1-E}{e^\frac{-t_1}{R_sC}}e^\frac{-t}{R_sC}
\end{equation}

In the (17) and (18), $v_c(t)$ and $i_c(t)$ are the voltage and current of capacitor respectively.
The value of $R_s$ is very small, therefore the time of capacitor charging is short and capacitor is charged rapidly. However, capacitor discharging state is a swinging relation. Thus, capacitor voltage can be adjusted to the desired value just by controlling the charging state of capacitor voltage.

\section{DETERMINATION OF CAPACITANCE OF FLYING CAPACITOR}
The capacitor charging and discharging states is shown in Fig. 7(b)-(d). The value of $R_s$ is very small, therefore the time of capacitor charging is short and capacitor’s capacitance determination is applied only for discharging state of capacitor.

\begin{equation}
i_c=c\frac{\Delta{V_c}}{\Delta{t}}
\end{equation}

Here, $\Delta{t}$ is the time of single switching duty cycle. $\Delta{V_c}$ is the capacitor rated voltage drop in one switching duty cycle which is assumed five percent of capacitor rated voltage. Consequently:

\begin{equation}
T_s=\Delta{t}
\end{equation}

\begin{equation}
\Delta{v_c}=\%5\frac{E}{2}
\end{equation}

Circuit current in Fig. 7, ($i_s$), is the sine wave that can be formulated as follow:

\begin{equation}
i_s=i_m\sin{(wt)}
\end{equation}

$i_c$ in (21) is equal to maximum value of current in (24), $I_m$, and therefore (25) is utilized to calculate the capacitor’s capacitance.

\begin{equation}
c=\frac{i_c\Delta{t}}{\Delta{v_c}}
\end{equation}

\section{MPPT IN THE PROPOSED INVERTER}
Maximum power of PV cells in any constant is shown by $P_{PV}$. The efficiency of the system is $\eta$, the output is $P_{ac}$. Therefore:

\begin{equation}
P_{ac}=\eta{P_{pv}}
\end{equation}

And:

\begin{equation}
P_{ac}=V_sI_s
\end{equation}

Here, $V_s$ and $I_s$ are the RMS value of grid voltage and current respectively. Therefore, if one assigns any values for $P_{PV}$ and $V_{s}$, the appropriate value for $I_{ref}$ capable of tracking $I_s$ determined and this guarantees the maximum output power from solar cells.

\section{VOLTAGE LOSS CALCULATION}
In this section, the loss calculation has been done for the proposed topology. The loss calculation contains switches losses and capacitor losses.

\subsection{Switches losses}
Mainly conduction and switching losses are calculated for switches. Every power switch contains a transistor (MOSFET) and a diode, and therefore, the instantaneous conduction loss of these elements is calculated using (28) and (29), respectively [21]:

\begin{equation}
P_{c,T}(t)=[R_Ti(t)]i(t)
\end{equation}

\begin{equation}
P_{c,D}(t)=[V_D+R_Di(t)]i(t)
\end{equation}

Where $V_D$ is considered as forward voltage drop of the diode. $R_T$ and $R_D$ are equivalent resistance of the transistor and diode. At any time $N_T(t)$ transistor and $N_D(t)$ diodes are in current path, so the average conduction power losses can be written as:

\begin{equation}
P_C=\frac{1}{2\pi}\int_{0}^{2\pi}[N_T(t)P_{c,T}(t)+N_D(t)P_{c,D}(t)]dt
\end{equation}

The switching losses occur during turn on and turn off period of switches. For simplicity, the linear approximation of voltage and current of switches during the switching period is considered. Based on this assumption, the switching loss is calculated as follows:

\begin{multline}
E_{off,J}=\int_{0}^{t_{off}}(v(t)i(t))dt= \\
\int_{0}^{t_{off}}[(\frac{V_{sw,J}t}{t_{off}})(-\frac{I''(t-t_{off}}{t_{off}})]dt=\frac{1}{6}V_{sw,J}I''t_{off}
\end{multline}

\begin{multline}
E_{on,J}=\int_{0}^{t_{on}}(v(t)i(t))dt= \\
\int_{0}^{t_{on}}[(\frac{V_{sw,J}t}{t_{on}})(-\frac{I'(t-t_{on}}{t_{on}})]dt=\frac{1}{6}V_{sw,J}I't_{on}
\end{multline}

$E_{off,J}$ and $E_{on,J}$ are turn off and turn on loss of the switch $J$, $I’’$ is the current through the switch before turning off, $I’$ is the current through the switches after turning on and $V_{sw,J}$ is the off-state voltage on the switch. The switching power loss is equal to the sum of all turns off and turns on energy losses in the fundamental frequency of the output voltage; as a result, average switching power loss can be calculated as follow:

\begin{equation}
P_{sw}=f\sum_{J=1}^{N_{switch}}(\sum_{k=1}^{N_{on,J}}{E_{on,JK}}+\sum_{k=1}^{N_{off,J}}{E_{off,JK}})
\end{equation}

$f$ is the fundamental frequency, $N_{on,J}$ and $N_{off,J}$ are the number of turning on and off the $J^{th}$ switch during fundamental frequency. $E_{on,JK}$ is the energy loss of the $J^{th}$ switch during the $k^{th}$ turning on and $E_{off,JK}$ is the energy loss of the $J^{th}$ switch during the $k^{th}$ turning off.

\subsection{Capacitor losses}
The capacitor losses consist of capacitor voltage ripple losses $P_{rip}$ and conduction losses $P_{cond}$. The voltage ripple of capacitor is calculated as follow:

\begin{equation}
\Delta{V_{rip}}=\frac{1}{C}\int_{t_-}^{t_+}i_Cdt
\end{equation}

Where $i_C$, the transient current of the capacitor, equals to the line current. $t_-$ to $t_+$ is the interval that the capacitor is connected in series with DC voltage sources. So, voltage ripple loss is written as follow:

\begin{equation}
P_{rip}=C\Delta{V_{rip}^2f}
\end{equation}

$f$ is the frequency of the output voltage. The conduction losses can be written as follow:

\begin{equation}
P_{cond}=2f\int_{t_-}^{t_+}r_Ci_C^2dt
\end{equation}

Where $r_c$ is the internal resistance of the capacitor. Total loss of the proposed topology is equal:

\begin{equation}
P_{loss}=P_C+P_{sw}+P_{rip}+P_{cond}
\end{equation}

Finally, the efficiency of converter is written as follow:

\begin{equation}
\eta=\frac{P_{out}}{P_{in}}=\frac{P_{in}-P_{loss}}{P_{in}}
\end{equation}

Where $P_{out}$ and $P_{in}$ are output power and input power of converter.

\section{SIMULATION RESULTS AND EXPERIMENTAL VERIFICATION}
The simulation has been done by PSCAD / EMTDC. The circuit used in the simulation and the experimental set up and the value of elements used in both of them are shown in Fig. 9 and Table III. Due to the limitation of laboratory components, DC voltage source is used instead of PV cells.

\begin{figure}[!t]
\centering
\captionsetup{justification=centering}
\includegraphics[width=3.5in]{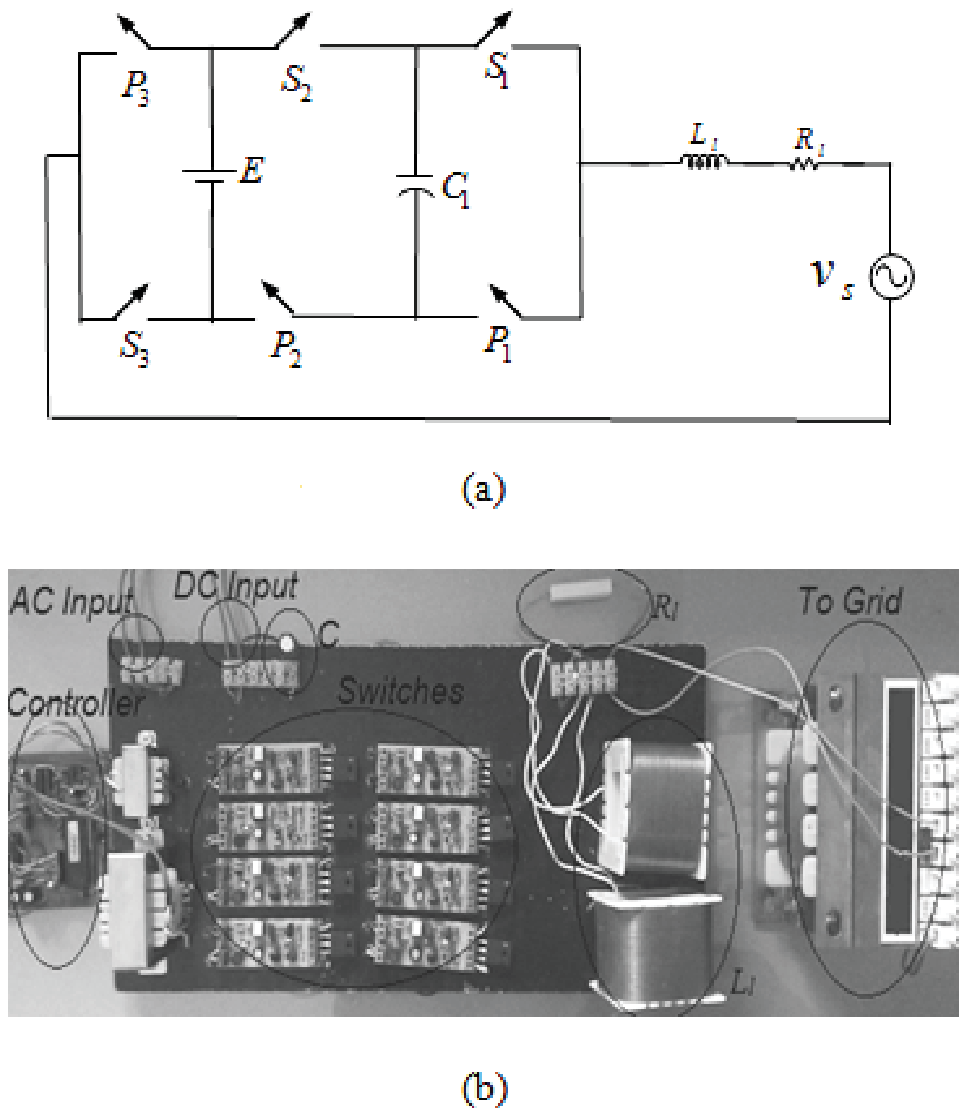}
\caption{(a) Circuit of simulation of 5-level proposed inverter and (b) Picture of experimental set up.}
\label{fig_sim}
\end{figure}

\begin{table}[!t]
\renewcommand{\arraystretch}{1.3}
\caption{Quantitative values used for simulation}
\label{Numerical values for the elements used in the simulation}
\centering
\begin{tabular}{|c||c||c|}
\hline
Parameter & Value & Attribute\\
\hline
$E$ & $60$ & Inverter Input Voltage\\
\hline
$V_m$ & $48$ & Grid maximum voltage\\
\hline
$R_l$ & $0.1$ & Line Resistance\\
\hline
$L_l$ & $0.005$ & Line Inductance\\
\hline
$C_1$ & $500$ & Flying Capacitor\\
\hline
$f_s$ & $10$ & Switching Frequency\\
\hline
$f_g$ & $50$ & Grid Frequency\\
\hline
$R_{D-S,on}$ & $0.4$ & Transistor Drain-Source On-State Resistance\\
\hline
\end{tabular}
\end{table}

For further analysis of inverter’s performance, input DC voltage source is changing from 60 V to 65 V in 0.04 s and from 65V to 55 V in 0.06 s, the results are shown in Fig. 10. According to Fig. 10, the proposed inverter with proposed control method produces 5-level voltage and tracks the reference current properly in the variation of DC source. Also, Flying capacitor’s voltage is balanced in its specified value (half of the input DC source).

\begin{figure}[!t]
\centering
\captionsetup{justification=centering}
\includegraphics[width=3.5in]{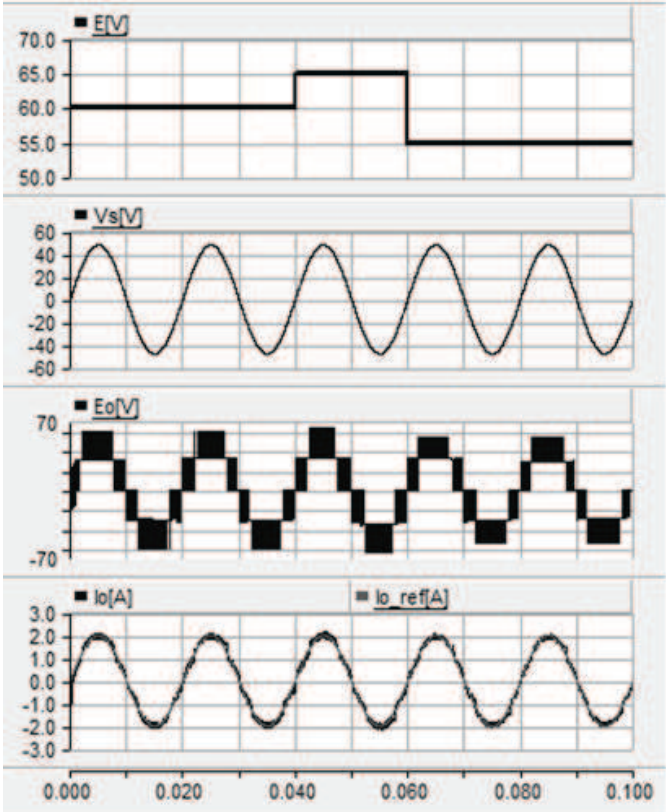}
\caption{Input DC source (E), grid voltage (Vs), inverter output voltage (Eo) and inverter output current (Io).}
\label{fig_sim}
\end{figure}

In the experimental circuit, inverter output voltage and current for 5-level output voltage is shown in Fig. 11. This figure indicates that the proposed inverter produces an output voltage waveform with 5 levels with maximum output level of 60 V. Also in this inverter, output current of approximately 2 A tracks the reference current properly that is in phase with grid voltage, therefore the proposed inverter can exchange active power with the grid. In the laboratory condition the measured output and input power are about 48 (W) and 52.8 (W), respectively. Therefore, the efficiency is equal 90.91\%. Based on the efficiency calculation, it is about 91.13\%. Accordingly, the computed efficiency has good accordance with the measured efficiency. Voltage and voltage ripples of flying capacitor are shown in Fig. 12. Base on Fig. 12, the voltage of flying capacitor is balanced on the desired value with 0.2 ripples.

\begin{figure}[!t]
\centering
\captionsetup{justification=centering}
\includegraphics[width=3.5in]{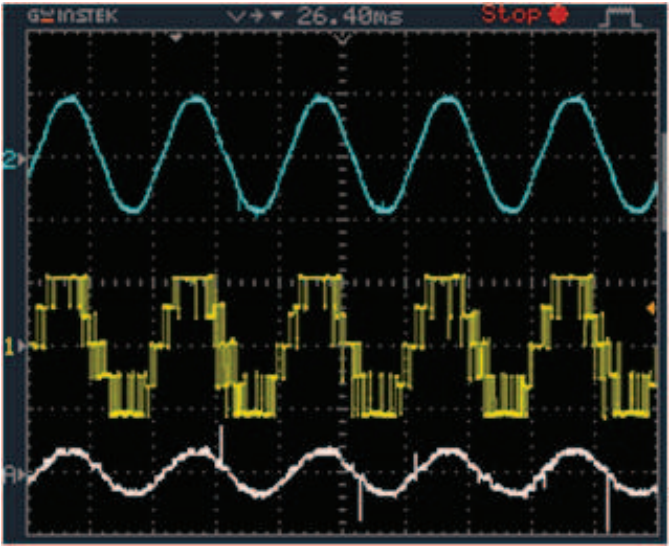}
\caption{Grid voltage (50V/div), inverter output voltage (50V/div) and current (4A/div), 10ms/div.}
\label{fig_sim}
\end{figure}

\begin{figure}[!t]
\centering
\captionsetup{justification=centering}
\includegraphics[width=3.5in]{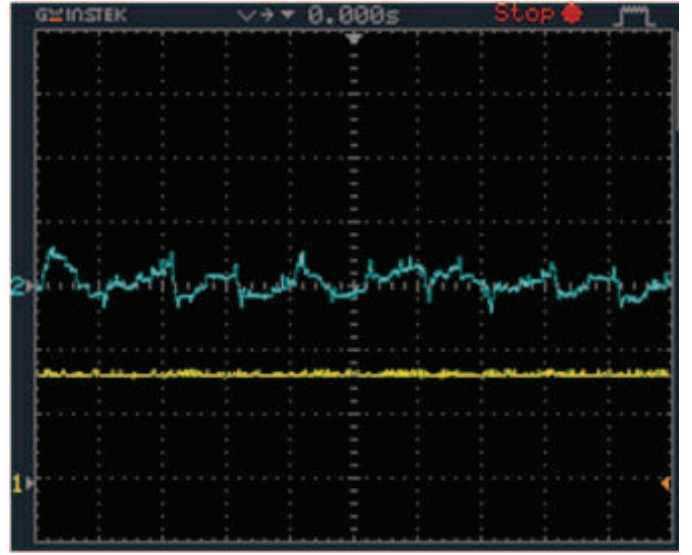}
\caption{Voltage ripples (200mV/div) and voltage of flying capacitor (20V/div), 5ms/div.}
\label{fig_sim}
\end{figure}

\section{Conclusion}
In the proposed inverter compared to the structure of DFCM, independent of the inverter cell number and the number of generated voltage levels, by replacing only two unidirectional switches with two bidirectional switches, network connectivity with high safety is provided. Also, this inverter with proposed control method has unique abilities such as capacitor voltage balancing in any desired value and obtaining maximum power from PV cells. Theoretically there is no limitation for the number of flying capacitors for balancing the voltage. The proposed control method tracks maximum power point of PV cells through assuring the equal magnitude for the input and output power. Through this unique controlling methodology, the MPPT is performed just by the inverter itself. Finally, the mathematical analysis and simulation results are used to verify the operation of proposed inverter and control method.
The design, engineering and optimization procedure explained in this work provides further insight for designing multilevel inverters to be used in hybrid systems equipped with various energy storage devices (i.e. redox flow batteries and regenerative hydrogen fuel cell systems). 
\begin{table}[!t]
\renewcommand{\arraystretch}{1.3}
\caption{Abbreviations}
\label{LIST OF ABBREVIATIONS AND SYMBOLS}
\centering
\begin{tabular}{|c||c|}
\hline
DC & Direct Current\\
\hline
MPPT & Maximum Power Point Tracking\\
\hline
NPC & Neutral Point Clamped\\
\hline
CM & Cascaded Multicell\\
\hline
FCM  & Flying Capacitor Multicell\\
\hline
SM & Stacked Multicell\\
\hline
DFCM  & Double Flying Capacitor Multicell\\
\hline
PV & Photovoltaic\\
\hline
THD & Total Harmonic Distortion\\
\hline
MOSFET & Metal–Oxide–Semiconductor Field-Effect Transistor\\
\hline
\end{tabular}
\end{table}

\begin{table}[!t]
\renewcommand{\arraystretch}{1.3}
\caption{Symbols}
\label{LIST OF ABBREVIATIONS AND SYMBOLS}
\centering
\begin{tabular}{|c||c|}
\hline
$n$ & Number of cells\\
\hline
$E$ & Magnitude of DC voltage source\\
\hline
$v_s$ & Grid voltage\\
\hline
$i_r$ & Reference current\\
\hline
$V_m$ & Grid maximum voltage\\
\hline
$i_s$ & Grid current\\
\hline
$\omega$ & Angular frequency of grid voltage\\
\hline
$R_s$ & Resistance of switches per current path loop\\
\hline
$N_T(t)$ & Number of Transistor in current path at any time\\
\hline
$N_D(t)$ & Number of Diode in current path at any time\\
\hline
$r_C$ & Internal resistance of capacitor\\
\hline
$f_g$ & Grid frequency\\
\hline
$f_s$ & Switching frequency\\
\hline
\end{tabular}
\end{table}




\ifCLASSOPTIONcaptionsoff
  \newpage
\fi

\bibliographystyle{ieeetran} 









\end{document}